\documentclass[12pt]{article}
\usepackage{graphicx}
\usepackage{slashed}
\DeclareGraphicsExtensions{.pdf}
\usepackage{float} 
\usepackage{amsmath}
\usepackage{amsfonts}   
\usepackage{amssymb}

\begin{document}
\date{}
\title{{\bf{\Large On anisotropic black branes with Lifshitz scaling}}}
\author{
 {\bf {\normalsize Dibakar Roychowdhury}$
$\thanks{E-mail:  dibakarphys@gmail.com, dibakarr@iitk.ac.in}}\\
 {\normalsize  Indian Institute of Technology, Department of Physics,}\\
  {\normalsize Kanpur 208016, Uttar Pradesh, India}
}

\maketitle
\begin{abstract}
In this paper, based on the method of scalar perturbations, we construct the \textit{anisotropic} charged Lifshitz background perturbatively upto leading order in the anisotropy. We perform our analysis both in the extremal as well as in the non extremal limit. Finally, we probe the so called superfluid phase of the boundary theory and explore the effects of anisotropy on the superconducting condensate.
\end{abstract}

\section{Overview and Motivation}

For the past couple of years, the \textit{holographic} constructions of anisotropic black black branes in an asymptotically anti-de Sitter (AdS) space time has been an active field of investigation due to its various remarkable implications that ranges between the physics of QCD to various exotic phases in usual condensed matter applications \cite{Mateos:2011tv}-\cite{Cheng:2014sxa}. Motivated from these analysis, recently the authors in \cite{Iizuka:2012wt} have constructed four dimensional gravitational theories (with $ AdS_{4} $ asymptotic) those are homogeneous but do not preserve the usual $ SO(2) $ rotational invariance along the two dimensional spatial hypersurface. In particular, motivated by the analysis performed in \cite{Mateos:2011tv}-\cite{Mateos:2011ix}, the authors in their analysis \cite{Iizuka:2012wt} have deformed the usual charged $ AdS_{4} $ black brane configuration by a relevant scalar deformation that depends linearly on one of the spatial directions of the brane. Such deformations turned out to be the true source of anisotropy in the sense that they generate the anisotropic energy momentum tensor in which, $ T^{x}\ _{x}\neq  T^{y}\ _{y}$. In their analysis \cite{Iizuka:2012wt}, the authors have considered the examples from Einstein-Maxwell-Dilaton theory and Einstein-Maxwell-Dilaton-Axion theory with $ SL(2,R) $ symmetry where these scalar deformations could be thought of as perturbations either due to the dilation or that of the axion profile in the bulk. 

Motivated from these analysis, the purpose of the present exercise is to extend the notion of anisotropy from AdS black branes to black brane configurations with asymptotic Lifshitz scaling \cite{Kachru:2008yh}-\cite{Taylor:2008tg} that does not seem to have addressed in the literature so far. Before we proceed further, two issues are extremely important to understand. First of all, it is important to realize that gravitational constructions over (asymptotic) Lifshitz like geometries (those are dual descriptions of Quantum Field Theories at Lifshitz fixed points) play crucial role in various condensed matter applications\footnote{Particularly, in this context it is noteworthy to mention that in nature there exists several ferromagnetic materials like, $ MnSi $ and some unconventional cuprate superconductors as well as iron pnictides, electronic nematics like $ Sr_{3}Ru_{2}O_{7} $ which exhibit Lifshitz type anisotropic scaling (namely, $ t \rightarrow \lambda^{z}t $ and $ \textbf{x}\rightarrow \lambda \textbf{x} $, $ z>1 $) near its quantum critical point \cite{Iqbal:2011ae}.}, for example, the hydrodynamic description of quantum (non Fermi) liquids near the quantum critical point \cite{Hoyos:2013eza} and in particular explaining the low temperature anomalous behavior of the specific heat in heavy Fermion compounds like $ CeCu_{5.9}Au_{0.1} $ for which the value of the dynamic critical exponent turns out to be, $ z\approx 2.7 $ \cite{Brynjolfsson:2010rx}. It is in fact quite evident by now that holographic models with anisotropic (Lifshitz like) scaling are the basic building blocks for theories consisting de-confined compressible fermionic excitation distributed over hidden Fermi surfaces \cite{Ogawa:2011bz}-\cite{Huijse:2011ef}. All the above discussions indeed suggest that holographic descriptions of Lifshitz like geometries play extremely important role in various condensed matter applications where the usual perturbative approach to field theories does not work well.

The second important issue turns out to be the role of anisotropy in holographic theories comprising (asymptotic) Lifshitz isometry group. In order to address this issue, we need to understand the following: In usual condensed matter systems, due to the presence of background lattice and/or crystal structures one could in principle imagine the breaking of rotational symmetry associated with the Fermi surfaces. Therefore, under these circumstances, holographic constructions (those are homogeneous but not isotropic) with Lifshitz asymptotic might play a crucial role in order to probe heavy Fermion compounds near its (anisotropic) Fermi surfaces. The present analysis could be regarded as the first step towards this direction.

The organization of the paper is the following: In Section 2, we review the essential features of \textit{unperturbed} (original) charged Lifshitz backgrounds those are originally discussed in \cite{Tarrio:2011de}. In Section 3, we cook up corresponding anisotropic Lifshitz background in the non extremal limit. In Section 4, we discuss the corresponding extremal limit. In Section 5, as an application of our holographic construction, we probe the superfluid phase of the boundary theory and explore the effects of anisotropy on the superconducting condensate. Finally, we conclude in Section 6.

\section{The gravity set up}
We start our analysis with a formal introduction to the gravitational set up in the bulk. It is to be noted that the ultimate goal of our analysis is to construct the anisotropic charged asymptotically Lifshitz black brane solutions in ($ 3+1 $) dimensions. To do that we first review the unperturbed \textit{homogeneous} as well as the \textit{isotropic} charged Lifshitz black brane solutions in ($ 3+1 $) dimensions. The theory that we describe in the bulk essentially consists of two $ U(1) $ gauge fields ($ \mathcal{A}_{i} $) and a dilaton field $ \varphi $ coupled to four dimensional Einstein's gravity in the presence of a negative cosmological constant ($ \Lambda $). The scalar (dilaton) field ($ \varphi $) and one of the abelian one forms ($ \mathcal{A}_{1} $) are required in order to support an asymptotically Lifshitz space time at the boundary. While on the other hand, the black brane is supposed to be charged under the second $ U(1) $ field ($  \mathcal{A}_{2}  $). The resulting diffeomorphism invariant action in ($ 3+1 $) dimensions could be formally expressed as \cite{Tarrio:2011de}, 
\begin{eqnarray}
S=-\frac{1}{16 \pi G_{4}}\int d^{4}x\sqrt{-g} \left(R-2\Lambda -\frac{1}{2}(\partial \varphi)^{2}-\frac{1}{4}\sum_{i=1}^{2}\exp (\lambda_{i}\varphi) \mathsf{F}_{i}^{2}\right).
\label{E1} 
\end{eqnarray} 

The equations of motion that readily follow from the above action (\ref{E1}) could be formally expressed as \cite{Tarrio:2011de},
\begin{eqnarray}
R_{ab}-\Lambda g_{ab}&=&\frac{1}{2}\left(\partial_{a}\varphi \partial_{b}\varphi \right)+\frac{1}{2}\sum_{i=1}^{2}\exp(\lambda_{i}\varphi) \left( \mathsf{F}^{c}\ _{ia}\mathsf{F}_{icb}-\frac{g_{ab}}{4}\mathsf{F}_{i}^{2}\right)=\mathsf{T}_{ab}\nonumber\\
\nabla^{2}\varphi &=&\frac{1}{4}\sum_{i=1}^{2}\lambda_{i}\exp (\lambda_{i}\varphi)\mathsf{F}_{i}^{2}\nonumber\\
0&=&\nabla_{a}(\exp(\lambda_{i}\varphi)\mathsf{F}^{ab}_{i}). 
\label{E3}
\end{eqnarray}
The resulting $ U(1) $ charged Lifshitz black brane configuration turns out to be\footnote{From now on we set $ L=1 $.},
\begin{eqnarray}
ds^{(0)2}&=&-\left( \frac{r_H}{u}\right)^{2z}f^{(0)}(u)dt^{2}+\frac{du^{2}}{u^{2}f^{(0)}(u)}+\left(\frac{r_H}{u} \right)^{2}(dx^{2}+dy^{2})\nonumber\\
f^{(0)}(u)&=&1-\mathsf{M}\left(\frac{u}{r_H} \right)^{z+2} + \frac{\mathsf{Q}^{2}\mathfrak{a}^{-\sqrt{z-1}}}{4z}\left(\frac{u}{r_H} \right)^{2(z+1)}\nonumber\\
\mathsf{F}_{1ut}&=&\sqrt{2(z-1)(z+2)}\mathfrak{a}^{\sqrt{\frac{1}{(z-1)}}}\frac{r_H^{z+1}}{u^{z+1}},~~
\mathsf{F}_{2ut}=\mathsf{Q}\mathfrak{a}^{-\sqrt{z-1}}\frac{u^{z+1}}{r_H^{z+1}}\nonumber\\
e^{\varphi} &=& \mathfrak{a}\left( \frac{r_H}{u}\right)^{2\sqrt{(z-1)}},~~\Lambda = -\frac{(z+2)(z+1)}{2}\nonumber\\
\lambda_{1}&=&-\frac{2}{\sqrt{z-1}},~~\lambda_{2}=\sqrt{z-1}
\label{E4}  
\end{eqnarray}
where, $ \mathsf{M} $ is a constant that could be related to the mass of the black brane and $ \mathsf{Q} $ is the $ U(1) $ charge of the black brane. Note that here $ u(=r_H/r) $ could be thought of as the (inverse) radial coordinate such that the horizon of the black brane is located at $ u=1 $. On the other hand, the boundary of the space time is realized in the limit $ u \rightarrow 0 $. At this stage it is noteworthy to mention that the above solution (\ref{E4}) indeed represents a homogeneous as well as isotropic black brane configuration in ($ 3+1 $) dimensions. By isotropy we always mean the $ SO(2) $ rotational invariance along the $ (x-y) $ directions of the brane and the homogeneity implies the translational invariance over any constant $ u(=u_c) $ hypersurface.  

Keeping the spirit of this intriguing fact, the purpose of our current analysis therefore would be to construct the corresponding homogeneous but \textit{anisotropic} charged Lifshitz black brane configurations in ($ 3+1 $) dimensions where the usual $ SO(2) $ rotational invariance is broken at any intermediate point during the holographic RG flow, while preserving the (asymptotic) Lifshitz symmetry algebra exactly at the UV fixed point ($ u=0 $) of the boundary QFTs. Before we conclude our discussion of this section, it is important to emphasise that the anisotropy that we consider in our work eventually could be thought of as perturbations over the original homogeneous and isotropic charged Lifshitz background (\ref{E4}) where we will be mostly interested to explore the effects of anisotropy only upto leading order in the perturbation series.

\section{The anisotropic construction}
Keeping the spirit of our earlier discussions, we choose the following metric ansatz for the $ U(1) $ charged Lifshitz black brane configurations in ($ 3+1 $) dimensions namely,
\begin{eqnarray}
ds^{2}= -\left( \frac{r_H}{u}\right)^{2z}f(u)dt^{2}+\frac{du^{2}}{u^{2}f(u)}+e^{\mathcal{A}(u)+\mathcal{B}(u)}dx^{2}+e^{\mathcal{A}(u)-\mathcal{B}(u)}dy^{2}.
\label{E5}
\end{eqnarray}  
At this stage, it is noteworthy to mention that the above black brane configuration (\ref{E5}) is homogeneous at a fixed radial slice $ u=u_{c} $, whose generators are the usual generators of the Killing isometries namely, $ \partial_{x} $ and $ \partial_{y} $. On the other hand, due to the presence of the function $ \mathcal{B}(u) $, the above metric (\ref{E5}) suffers from the spatial anisotropies along the ($ x-y $) directions of the brane. The purpose of our present analysis would be to evaluate the function $ \mathcal{B}(u) $ analytically in a perturbation series considering the effects upto leading order in the anisotropy parameter. 

We first note down Einstein's equations (\ref{E3}) corresponding to the above anisotropic Lifshitz black brane configuration (\ref{E5}), which for the present case yield the set of following four equations namely,
\begin{eqnarray}
\frac{1}{2} r_H^{2 z} u^{-2 z} f \left(2 z f \left(z-u \mathcal{A}'\right)+u \left(\left(1-3 z+u \mathcal{A}'\right) f'+u f''\right)\right)&=&\mathsf{T}_{tt}+\Lambda g_{tt}\nonumber\\
f \left(2 z^2+2 u \mathcal{A}'+u^2 \mathcal{A}'^2+u^2 \mathcal{B}'^2+2 u^2 A''\right)+u \left(\left(1-3 z+u \mathcal{A}'\right) f'+u f''\right)
&=&-2u^{2}f(\mathsf{T}_{uu}+\Lambda g_{uu})\nonumber\\
u \left(u \left(\mathcal{A}'+\mathcal{B}'\right) f'+f \left(u \mathcal{A}'^2-(-1+z) \mathcal{B}'+\mathcal{A}'\left(1-z+u \mathcal{B}'\right)+u \left(\mathcal{A}''+\mathcal{B}''\right)\right)\right)
&=&-2e^{-(\mathcal{A}+\mathcal{B})}\mathsf{T}_{xx}-2\Lambda\nonumber\\
u \left(u \left(\mathcal{A}'-\mathcal{B}'\right) f'+f \left(u \mathcal{A}'^2+(-1+z) \mathcal{B}'-\mathcal{A}' \left(-1+z+u \mathcal{B}'\right)+u \left(\mathcal{A}''-\mathcal{B}''\right)\right)\right)
&=&-2e^{-\mathcal{A}+\mathcal{B}}\mathsf{T}_{yy}-2\Lambda.\nonumber\\
\end{eqnarray}

After some trivial algebra, the above set of equations could be re-arranged as,
\begin{eqnarray*}
\mathcal{B}''+\mathcal{B}'\left(\mathcal{A}'+\frac{1-z}{u}+\frac{f'}{f} \right)&=&-\frac{e^{-\mathcal{A}}}{u^{2}f}(e^{-\mathcal{B}}\mathsf{T}_{xx}-e^{\mathcal{B}}\mathsf{T}_{yy})\nonumber\\
\mathcal{A}''+\mathcal{A}'^{2}+\mathcal{A}'\left(\frac{f'}{f}+\frac{1-z}{u} \right)&=&-\frac{e^{-\mathcal{A}}}{u^{2}f}(e^{-\mathcal{B}}\mathsf{T}_{xx}+e^{\mathcal{B}}\mathsf{T}_{yy}) - 2\Lambda
\end{eqnarray*}
\begin{eqnarray}
f'' +\frac{f'}{u}(1-3z+u\mathcal{A}')+\frac{2z^{2}f}{u^{2}}-\frac{f(\mathcal{A}'^{2}-\mathcal{B}'^{2})}{(f+1)}
+\frac{2f\mathcal{A}'}{u(f+1)}\left(1-zf-u\left(\frac{f'}{f}+\frac{1-z}{u} \right)  \right)\nonumber\\
= \frac{2e^{-\mathcal{A}}}{u^{2}(f+1)}(e^{-\mathcal{B}}\mathsf{T}_{xx}+e^{\mathcal{B}}\mathsf{T}_{yy})+\frac{2f}{(f+1)}\left( \frac{u^{2(z-1)}}{f r_{H}^{2z}}(T_{tt}+\Lambda g_{tt})-(T_{uu}+\Lambda g_{uu})\right)+\frac{4f\Lambda}{(f+1)}. 
\label{E11}
\end{eqnarray}

It is now quite trivial to see that given the following ansatz for the dilaton ($ \varphi $) as well as the $ U(1) $ gauge fields namely, 
\begin{eqnarray}
\varphi &=& \varphi (u)\nonumber\\
\mathsf{A}_{ia}dx^{a}&=& \mathsf{A}_{it}(u)dt
\end{eqnarray}
the R.H.S. of the first equation of (\ref{E11}) vanishes identically,
\begin{eqnarray}
e^{-\mathcal{A}}(e^{-\mathcal{B}}\mathsf{T}_{xx}-e^{\mathcal{B}}\mathsf{T}_{yy})=0.
\end{eqnarray}

With this condition in hand, from the first equation in (\ref{E11}) we note that,
\begin{eqnarray}
\mathcal{B}'(u)= \mathsf{C}e^{-\int_{1}^{u}\left(\mathcal{A}'(u')+\frac{1-z}{u'}+\frac{f'(u')}{f(u')} \right)du' }
\end{eqnarray}
which is clearly undefined as in the near horizon ($ u'=1 $) limit the integrand blows up. If we demand that the metric is regular everywhere in the space time, then we are forced to set, $ \mathsf{C}=0 $, which eventually suggests that $ \mathcal{B} $ does not change along the radial coordinate ($ u $). In other words, $ \mathcal{B} $ does not flow along the radial coordinates and therefore it is same at any given radial slice, $ u=u_c $. Now, given the initial data over the hypersurface $ u=u_c $, one can always generate coordinate transformations in the $ (x-y) $ plane such that the metric is diagonal at that hypersurface namely, $ \mathcal{B}(u_{c})=0 $. Combining both of these arguments together one essentially concludes that, $ \mathcal{B}(u)=0 $ which thereby implies the isotropy \cite{Iizuka:2012wt}. In other words, given the initial data (over some Cauchy surface) which is homogeneous as well as isotropic one always ends up with the homogeneous and the isotropic configurations during the (RG) flow along the radial coordinate ($ u $).  Similar observations could also be made in the context of pure Einstein's gravity without any matter content \cite{Iizuka:2012wt}. Therefore, in order to retain the anisotropy ($ \mathcal{B}\neq 0 $), one must have a non zero contribution on the R.H.S. of the first equation in (\ref{E11}) which eventually implies that the stress tensor must satisfy the following criterion namely \cite{Iizuka:2012wt},
\begin{eqnarray}
T^{x}\ _{x}\neq T^{y}\ _{y}.
\end{eqnarray}

Therefore, in order to generate the homogeneous and the \textit{anisotopic} solution we invite an additional scalar perturbation in our theory namely\footnote{Our choice of $ \Phi (x)$ is based on the physical fact that in the dual field theory sector we have deformed the theory by adding a relevant deformation ($ \Theta(x) $) which drives the system away from its (UV) Lifshitz fixed point and depends linearly on one of the spatial coordinates ($ x $) of the brane. In the framework of Gauge/gravity duality, the (scalar) operator $ \Theta(x) $ is dual to $\Phi (x) $ in the bulk. Therefore, the entire set up eventually suggests that the isotropy of the dual QFT is broken as a consequence of the anisotropically extended branes in the bulk. Finally, in the limit $ \mathfrak{K} \rightarrow 0 $, one should be able to recover the usual (isotropic) Lifshitz black brane configurations (\ref{E4}) in ($ 3+1 $) dimensions.}\cite{Iizuka:2012wt},
\begin{eqnarray}
\Phi(x)=\mathfrak{K} x
\label{E12}
\end{eqnarray}
which thereby modifies our starting action as,
\begin{eqnarray}
S=-\frac{1}{16 \pi G_{4}}\int d^{4}x\sqrt{-g} \left(R-2\Lambda -\frac{1}{2}(\partial \varphi)^{2}-\frac{1}{2}(\partial \Phi)^{2}-\frac{1}{4}\sum_{i=1}^{2}\exp (\lambda_{i}\varphi) \mathsf{F}_{i}^{2}\right).
\label{e13} 
\end{eqnarray}
Here $ \mathfrak{K} $ is the so called anisotropy parameter that could be considered to be infinitesimally small. The modified stress tensor that directly follows from (\ref{e13}) could be formally expressed as,
\begin{eqnarray}
T^{a}\ _{b}=\frac{1}{2}\partial^{a}\varphi \partial_{b}\varphi +\frac{1}{2}\partial^{a}\Phi (x) \partial_{b}\Phi (x) + \frac{1}{2}\sum_{i=1}^{2}\exp(\lambda_{i}\varphi) \left( \mathsf{F}^{ca}\ _{i}\mathsf{F}_{icb}-\frac{\delta^{a}_{b}}{4}\mathsf{F}_{i}^{2}\right)
\label{e14}
\end{eqnarray}
which finally yields,
\begin{eqnarray}
T^{x}\ _{x}- T^{y}\ _{y}=\frac{\mathfrak{K}^{2}}{2}e^{-(\mathcal{A}+\mathcal{B})}.
\label{e15} 
\end{eqnarray}

Since the first non trivial effect of anisotropy appears at the quadratic level in $ \mathfrak{K} $ (\ref{e15}), therefore for the present purpose our analysis it is indeed sufficient to solve the background (\ref{E5}) perturbatively upto quadratic order in $ \mathfrak{K} $. One should also take a note on the fact that since the metric does not depend on the spatial coordinate $ x $, therefore the background is still homogeneous on a given hypersurface, $ u=u_{c} $. Finally, it is noteworthy to mention that the above ansatz (\ref{E12}) trivially satisfies the equation of motion for $ \Phi (x) $ itself,
\begin{eqnarray}
\nabla^{2}\Phi = 0.
\end{eqnarray}

Our next task, therefore would be to solve the above set of equations (\ref{E11}) perturbatively in the (anisotropic) parameter $ \mathfrak{K} $. To do that, we expand solutions corresponding to (\ref{E11}) perturbatively around the original (isotropic) Lifshitz background (\ref{E4}) as,
\begin{eqnarray}
\mathcal{A}&=& \mathcal{A}^{(0)}+\mathfrak{K}^{2}\mathcal{A}^{(1)}+ \mathcal{O}(\mathfrak{K}^{4})\nonumber\\
\mathcal{B}&=& \mathfrak{K}^{2}\mathcal{B}^{(1)}+\mathcal{O}(\mathfrak{K}^{4})\nonumber\\
f&=& f^{(0)}+\mathfrak{K}^{2}f^{(1)}+\mathcal{O}(\mathfrak{K}^{4})
\label{E20}
\end{eqnarray}
where we ignore all the effects of anisotropy beyond the quadratic order in $ \mathfrak{K} $. The corresponding expression for the function $ \mathcal{A}^{(0)} $ turns out to be,
\begin{eqnarray}
\mathcal{A}^{(0)}(u)=2 \log \left(\frac{r_H}{u} \right). 
\label{eq17}
\end{eqnarray}

With (\ref{eq17}) in hand, from (\ref{E3}) it is in fact quite straightforward to show,
\begin{eqnarray}
e^{\lambda_{i}\varphi}\mathsf{F}_{iut}\approx \frac{\mathsf{G}}{r_H^{2-z}u^{z-1}}(1-\mathfrak{K}^{2}\mathcal{A}^{(1)})
\end{eqnarray}
which clearly suggests that in the asymptotic limit the zeroth order solutions dominate.

Computing all the components of the stress tensor (\ref{e14}) and plugging it back into (\ref{E11}), we finally arrive at the following set of equations,
\begin{eqnarray*}
\mathcal{B}''+\mathcal{B}'\left(\mathcal{A}'+\frac{1-z}{u}+\frac{f'}{f} \right)+\mathfrak{K}^{2}\frac{e^{-(\mathcal{A}+\mathcal{B})}}{2u^{2}f}&=&0\nonumber\\
\mathcal{A}''+\mathcal{A}'^{2}+\mathcal{A}'\left(\frac{f'}{f}+\frac{1-z}{u} \right)+2\Lambda +\mathfrak{K}^{2}\frac{e^{-(\mathcal{A}+\mathcal{B})}}{2u^{2}f}-\frac{1}{4u^{2}f}\sum_{i=1}^{2}\exp (\lambda_{i}\varphi) \mathsf{F}_{i}^{2}&=&0
\end{eqnarray*}
\begin{eqnarray}
f'' +\frac{f'}{u}(1-3z+u\mathcal{A}')+\frac{2z^{2}f}{u^{2}}-\frac{f(\mathcal{A}'^{2}-\mathcal{B}'^{2})}{(f+1)}
+\frac{2f\mathcal{A}'}{u(f+1)}\left(1-zf-u\left(\frac{f'}{f}+\frac{1-z}{u} \right)  \right)\nonumber\\
= \frac{\mathfrak{K}^{2} e^{-(\mathcal{A}+\mathcal{B})}}{u^{2}(f+1)}-\frac{2\Lambda}{u^{2}}+\frac{4f\Lambda}{(f+1)}-\frac{f \varphi'^{2}}{(f+1)}-\frac{(5+3f)}{4u^{2}(1+f)}\sum_{i=1}^{2}\exp (\lambda_{i}\varphi) \mathsf{F}_{i}^{2} . 
\label{E17}
\end{eqnarray}

As a next step of our analysis, we substitute (\ref{E20}) into (\ref{E17}) which eventually yields the set of three equations for the fluctuations at their leading order. In the following, we consider them one by one. We first consider the equation corresponding to $ \mathcal{B}^{(1)}(u) $ which for the present case turns out to be,
\begin{eqnarray}
\mathcal{B}''^{(1)}+\mathcal{B}'^{(1)}\left( \frac{f'^{(0)}}{f^{(0)}}-\frac{z+1}{u}\right) +\frac{1}{2r_H^{2}f^{(0)}}=0.
\label{E18}
\end{eqnarray}

The solution corresponding to (\ref{E18}) is indeed quite difficult to achieve for arbitrary values of the dynamic critical exponent ($ z $). To solve (\ref{E18}) and subsequently the other equations of motion, we therefore choose a specific value for the dynamic exponent namely, $ z=2 $. With this choice in hand, the solution corresponding to (\ref{E18}) turns out to be,
\begin{eqnarray}
\mathcal{B}^{(1)}(u)&=&\mathfrak{C}_{2}+\frac{u^2}{16 r_H^2}+\frac{e^{\frac{\mathsf{M} u^4}{r_H^4}} r_H^4\mathfrak{C}_{1}}{4\mathsf{M}}+\frac{e^{\frac{\mathsf{M} u^4}{r_H^4}} \sqrt{\pi } \text{Erf}\left[\frac{\sqrt{\mathsf{M}} u^2}{r_H^2}\right]}{32 \sqrt{\mathsf{M}}}.
\label{E19}
\end{eqnarray}
At this stage, it is to be noted that the above solution (\ref{E19}) is exact in the radial coordinate $ u $ and is valid for the entire range $ 1\leq  u<0 $. 

In the near boundary ($ u \rightarrow 0 $) limit, the above solution (\ref{E19}) turns out to be,
\begin{eqnarray}
\mathcal{B}^{(1)}(u) \approx  \left(\frac{r_H^4 \mathfrak{C}_{1}}{4 \mathsf{M}}+\mathfrak{C}_{2}\right)+\frac{u^2}{8 r_H^2}+\frac{1}{4} \mathfrak{C}_{1} u^4 +\mathcal{O}(u^{5}).
\end{eqnarray}

In order to determine the two unknown integration constants $ \mathfrak{C}_{1} $ and $ \mathfrak{C}_{2} $, we therefore need two boundary conditions. One of them is precisely the asymptotic boundary condition namely,
\begin{eqnarray}
\mathcal{B}^{(1)}(0)=0
\end{eqnarray}
which yields,
\begin{eqnarray}
\mathfrak{C}_{2}=-\frac{r_H^4 \mathfrak{C}_{1}}{4 \mathsf{M}}.
\label{E21}
\end{eqnarray}

On the other hand, expanding the function (\ref{E19}) around $ u=1 $ we obtain,
\begin{eqnarray}
\mathcal{B}^{(1)}(u)\approx \frac{1}{32} \left(\frac{2}{r_H^2}+32 \mathfrak{C}_{2}+\frac{e^{\frac{\mathsf{M}}{r_H^4}} \left(8 r_H^4 \mathfrak{C}_{1}+\sqrt{\mathsf{M}} \sqrt{\pi } \text{Erf}\left[\frac{\sqrt{\mathsf{M}}}{r_H^2}\right]\right)}{\mathsf{M}}\right)\nonumber\\
+\frac{1}{8 r_H^4}\left(2 r_H^2+e^{\frac{\mathsf{M}}{r_H^4}} \left(8r_H^4 \mathfrak{C}_{1}+\sqrt{\mathsf{M}} \sqrt{\pi } \text{Erf}\left[\frac{\sqrt{\mathsf{M}}}{r_H^2}\right]\right)\right) (u-1)+\mathcal{O}(u-1)^{2}.
\label{E22}
\end{eqnarray}

Furthermore from (\ref{E18}) we note that,
\begin{eqnarray}
\mathcal{B}^{'(1)}(u)|_{u \rightarrow 1}\approx \frac{2 \mathfrak{a} r_H^4}{16 \mathfrak{a}\mathsf{M} r_H^2-3 \mathsf{Q}^2}+\mathcal{O}(u-1).
\label{E23}
\end{eqnarray}

Finally, using (\ref{E22}) and (\ref{E23}) one of the coefficients turns out to be,
\begin{eqnarray}
\mathfrak{C}_{1} &=&\frac{1}{8r_H^{4}}\left[\left(\frac{16 \mathfrak{a} r_H^8}{16 \mathfrak{a}\mathsf{M} r_H^2-3 \mathsf{Q}^2}-2r_H^{2} \right)e^{-\frac{\mathsf{M}}{r_H^4}}  -\sqrt{\mathsf{M}} \sqrt{\pi } \text{Erf}\left[\frac{\sqrt{\mathsf{M}}}{r_H^2}\right]\right]
\end{eqnarray}
which trivially fixes the other coefficient also.

Next, we note down the equation corresponding to the fluctuation $\mathcal{A}^{(1)} $ which for the present case turns out to be,
\begin{eqnarray}
\mathcal{A}''^{(1)}+\left( \frac{f'^{(0)}}{f^{(0)}}-\frac{5}{u}\right) \mathcal{A}'^{(1)}-\frac{2}{u f^{2(0)}}(f^{(0)}f'^{(1)}-f'^{(0)}f^{(1)})-\frac{u^{4} f^{(1)}}{4r_H^{4}f^{2(0)}}\sum_{i=1}^{2}e^{\lambda_{i}\varphi} \mathsf{F}_{iut}^{2}+\frac{1}{2r_H^{2}f^{(0)}}=0.
\label{E25}
\end{eqnarray}
It is indeed quite difficult to solve the above equation (\ref{E25}) for any generic value of $ u $. However, for the purpose of our present analysis, it is in fact sufficient to extract out the boundary behaviour of $ \mathcal{A}^{(1)}(u) $. Taking the $ u \rightarrow 0 $ limit of (\ref{E25}) and considering the asymptotic Lifshiz symmetry (which in turn sets, $ f^{(1)}(0)= f^{'(1)}(0)\sim 0$) we find,
\begin{eqnarray}
\mathcal{A}^{(1)}(u)\sim \frac{u^2}{16 r_H^2} + \mathcal{O}(u^{5})
\end{eqnarray}
which thereby clearly vanishes at the boundary in order to preserve the asymptotic Lifshitz symmetry algebra. Similar observations have been made earlier in the context of asymptotic $ AdS $ spaces \cite{Iizuka:2012wt}. 

Finally, we consider the fluctuations $ f^{(1)}(u) $ in the metric and solve them in the asymptotic limit. In the asymptotic ($ u \rightarrow 0 $) limit, the equation corresponding to $ f^{(1)}(u) $ turns out to be,
\begin{eqnarray}
f^{''(1)}-\frac{5}{u}f^{'(1)} +\left( \frac{10}{u^2}+6\right) f^{(1)}-\frac{1}{2r_H^{2}}\approx 0
\label{e29}
\end{eqnarray}
whose solution could be formally expressed as,
\begin{eqnarray}
f^{(1)}(u)\approx \frac{u^2(1-3u^{2})}{4 r_H^2}+\mathcal{O}(u^{6})
\end{eqnarray}
which satisfies (\ref{e29}) completely upto quadratic order in the variable $ u $.
 
Collecting all these fluctuations together, below we enumerate all the different components of the (anisotropic) metric that preserve the asymptotic ($ z=2 $) Lifshitz symmetry algebra,
\begin{eqnarray}
g_{tt}(u)&=&-\left( \frac{r_H}{u}\right)^{4} \left(f^{(0)}(u)+\frac{\mathfrak{K}^{2}u^{2}(1-3u^{2})}{4 r_H^{2}} \right) +\mathcal{O}(\mathfrak{K}^{4})\nonumber\\
g_{uu}(u)&=&\frac{1}{u^{2}f^{(0)}}\left(1-\frac{\mathfrak{K}^{2}u^{2}(1-3u^{2})}{4r_H^{2}f^{(0)}} \right) +\mathcal{O}(\mathfrak{K}^{4})\nonumber\\
g_{xx}(u)&=&\left( \frac{r_H}{u}\right)^{2}\left(1+\mathfrak{K}^{2}\left(  \frac{3u^2}{16 r_H^2}+\frac{1}{4} \mathfrak{C}_{1} u^4 \right)  \right)+\mathcal{O}(\mathfrak{K}^{4})\nonumber\\
g_{yy}(u)&=& \left( \frac{r_H}{u}\right)^{2}\left(1-\mathfrak{K}^{2}\left(  \frac{u^2}{16 r_H^2}+\frac{1}{4} \mathfrak{C}_{1} u^4 \right)  \right)+\mathcal{O}(\mathfrak{K}^{4}).
\label{E29}
\end{eqnarray}
\section{The extremal limit}
Before we conclude this section, it would be nice to have some qualitative understanding of our analysis near the extremal limit of the Lifshitz black brane (\ref{E5}). To do that, we first note down the Hawking temperature corresponding to the unperturbed ($ z=2 $) Lifshitz black brane configuration (\ref{E4}),
\begin{eqnarray}
T = \frac{1}{16 \pi}\frac{\left(\mathsf{Q}^2-16 \mathfrak{a}r_H^6\right)}{\mathfrak{a}r_H^{5}}.
\label{E30}
\end{eqnarray}

From (\ref{E30}), we note that the extremal limit eventually corresponds to,
\begin{eqnarray}
\mathsf{Q}^2=16 \mathfrak{a}r_H^6.
\label{E31}
\end{eqnarray}

Substituting (\ref{E31}) into the unperturbed metric (\ref{E4}), the near horizon structure of the extremal Lifshitz black brane configuration turns out to be,
\begin{eqnarray}
f^{(0)}(u)=12 (1-u)^{2}+\mathcal{O}((1-u)^{3})
\label{E32}
\end{eqnarray}
which thereby exhibits a double pole structure as that of the usual extremal $ AdS $ black holes \cite{Iizuka:2012wt}. With the above result (\ref{E32}) in hand, from (\ref{E18}) it is indeed quite trivial to check,
\begin{eqnarray}
\mathcal{B}^{(1)}(u)\sim \log |1-u|
\end{eqnarray}
which therefore clearly exhibits a logarithmic divergence in the near horizon limit of the extremal Lifshitz black brane. The nature of this divergence is indeed identical to that of the usual extremal $ AdS $ black branes \cite{Iizuka:2012wt}. The upshot of the above analysis could therefore be framed as follows: Like in the usual $ AdS $ scenario \cite{Iizuka:2012wt}, the whole perturbative treatment of anisotropy eventually breaks down in the near extremal limit of the Lifshitz black brane configuration (\ref{E5}). The key reason for this lies on the fact that as the temperature gets lower the effects of higher curvature corrections to the metric become important and as a result these higher curvature effects near the horizon of the extremal brane becomes significantly large which eventually take us away from the usual notion of two derivative Einstein's gravity. 

Before we conclude this section, a few important remarks are in order. First of all, in order to check the stability of the new set of (anisotropic) Lifshitz solutions (\ref{E29}), one should in principle study the quasi normal modes over this newly constructed anisotropic (Lifshitz) background (\ref{E29}). One way to do that is to turn on small (hydrodynamic) fluctuations over (\ref{E29}) and study various response parameters of the boundary hydrodynamics. We (partially) address some of these issues in the subsequent sections. Finally, and most importantly, the anisotropic ansatz (\ref{E5}) is not the most general one to start with. In particular, one should be able to note that the present choice of our scalar field (\ref{E12}) eventually yields, $ T^{y}\ _{x}=0 $, which thereby fixes, $ g_{xy} =0$. Therefore, depending on the matter content of the theory, one could in principle have off diagonal metric components as well which would be a much more general set up to look for.

\section{One application: AdS/CMT}
As an application of our holographic techniques, in the present section we study the phenomenon of superconductivity\footnote{Actually it is a superfluid phase rather than the superconductor. The reason for this is that the condensation of the scalar hair ($ \psi $) in the bulk essentially breaks the \textit{global} $ U(1) $ symmetry of the boundary theory\cite{Hartnoll:2008vx}. Moreover, the UV fixed point of the sperfluid phase that we describe here is Lifshitz scale invariant and not the usual conformal invariant.} in the so called \textit{probe} approximation over the fixed anisotropic background (\ref{E5}). To start with, we consider the following action \cite{Hartnoll:2008vx},
\begin{eqnarray}
S=\int d^{4}x \left(-\frac{1}{4}\mathfrak{F}^{2}-|D_{a}\psi|^{2} -m^{2}|\psi|^{2}\right) \label{E72}
\end{eqnarray} 
in the\textit{ probe} limit over the fixed background (\ref{E5}) where, $ D_{a} $ stands for the usual gauge covariant derivative. Before we proceed further, it is customary to note down the Hawking temperature ($ T $) corresponding to the above black brane configuration (\ref{E5}),
\begin{eqnarray}
T =\frac{r_H^{z-1}}{4\pi}f'(1).
\end{eqnarray}

To start with, we consider the following ansatz for the $ U(1) $ gauge field and the scalar field namely \cite{Hartnoll:2008vx},
\begin{eqnarray}
\mathfrak{A}_{a}&=&(\phi (u),0,0,0)\nonumber\\
\psi &=& \psi (u)
\end{eqnarray}
which satisfy the equations of motion of the following form,
\begin{eqnarray}
\phi'' +\left(\mathcal{A}'+\frac{z+1}{u} \right) \phi' -\frac{2 \phi \psi^{2}}{u^{2}f}&=&0\nonumber\\
\psi'' +\left(\frac{1-z}{u}+\mathcal{A}'+\frac{f'}{f} \right)\psi' +\frac{u^{2(z-1)}}{r_H^{2z}f^{2}}\phi^{2}\psi +\frac{2\psi}{u^{2}f} &=&0
\label{E75}
\end{eqnarray}
where, we have set, $ m^{2}=-2 $.

As a next step of our analysis, we consider the boundary conditions for the gauge field ($ \phi $) as well as the scalar field ($ \psi $). Regularity of the fields near the horizon ($ u=1 $) implies that,
\begin{eqnarray}
\phi (1)=0,~~\psi' (1)=-\frac{2\psi (1)}{f'(1)}.
\label{E76}
\end{eqnarray}
On the other hand, the asymptotic boundary conditions for these fields turn out to be,
\begin{eqnarray}
\phi (u)=\mu - \frac{\varrho}{r_H}\mathit{F} (u),~~\psi(u)= <\mathit{O}>\frac{u^{2}}{r_H^{2}}
\label{E77}
\end{eqnarray}
where, $ <\mathit{O}> $ is the vacuum expectation value of the scalar (condensation) operator for the boundary theory \cite{Hartnoll:2008vx} and $ \mathit{F} (u) $  is some asymptotic function whose detail structure could be determined from the asymptotic behaviour of the $ U(1) $ gauge field. 

With the above boundary conditions (\ref{E76}) and (\ref{E77}) in hand, we next proceed to solve the order parameter and/or the scalar condensation operator ($ <\mathit{O}> $) analytically as a perturbation in the anisotropy ($ \mathfrak{K}^{2} $). Substituting the asymptotic expansion of the scalar field, $ \psi^{2}\sim <\mathit{O}>^{2}u^{4} $ into the first equation of (\ref{E75}), the asymptotic structure of the equation corresponding to the $ U(1) $ gauge field (\ref{E75}) turns out to be,
\begin{eqnarray}
\phi'' +\left(\mathcal{A}'+\frac{3}{u} \right) \phi' \approx 0
\label{E78}
\end{eqnarray}
where, we have set $ z=2 $. The asymptotic solution corresponding to (\ref{E78}) could be formally expressed as,
\begin{equation}
\phi (u) \approx \mu -\frac{\varrho}{r_H}\left( \log u -\frac{\mathfrak{K}^2  u^2}{32 r_H^2}\right).
\label{E79}
\end{equation}

Before we proceed further, it is customary to point out the Taylor expansion of the gauge field ($ \phi (u) $) as well as the scalar field ($ \psi (u) $) near the horizon ($ u=1 $) of the black brane namely\cite{Gregory:2009fj}-\cite{Roychowdhury:2012hp},
\begin{eqnarray}
\phi (u)&=& \phi (1)-\phi'(1)(1-u)+\frac{1}{2}\phi''(1)(1-u)^{2}+..~..\nonumber\\
&=&-\phi'(1)(1-u)+\frac{1}{2}\phi''(1)(1-u)^{2}+..~..
\label{e89}
\end{eqnarray}
and,
\begin{eqnarray}
\psi(u)=\psi(1)+\psi'(1)(1-u)+\frac{1}{2}\psi''(1)(1-u)^{2}+..~..
\label{e90}
\end{eqnarray}
where, we choose $ \phi'(1)<0 $ and $ \psi'(1)>0 $ without loss of any generality \cite{Gregory:2009fj}-\cite{Roychowdhury:2012hp}.

Following the analytic techniques developed in \cite{Gregory:2009fj}, as a next step of our analysis, we first match the above asymptotic solution (\ref{E79}) to that with its near horizon ($ u \sim 1 $) expansion (\ref{e89}) of the gauge field ($ \phi(u) $) at any intermediate value of the radial variable $ u=u_{m} $ (say),
\begin{eqnarray}
-\zeta -\phi ''(1)(1-u_m)= -\frac{\varrho}{r_H}\left(\frac{1}{u_{m}} -\frac{\mathfrak{K}^2  u_{m}}{16 r_H^2}\right) 
 \label{E80}
\end{eqnarray}
where, we have defined $ \zeta =-\phi'(1) $. 

As a next step of our analysis, using the first equation of (\ref{E75}), one could further rewrite (\ref{E80}) as,
\begin{eqnarray}
\psi^{2}(1)\equiv \xi^{2}=\frac{f'(1)}{2} \left[\frac{1}{1-u_m}\left(\frac{\varrho}{\tilde{\zeta}u_m}\left(1-\frac{\mathfrak{K}^{2}u_m^{2}}{16 r_H^{2}} \right) -1 \right) -3-\mathcal{A}'(1)\right] 
\label{E81}
\end{eqnarray}
where we have replaced $ \phi''(1) $ in terms of $ \psi(1) $ and re-scaled, $ \tilde{\zeta}=\zeta r_H $.

Before we proceed further, it is customary to estimate the physical range of the matching parameter $ u_m $. Considering the fact, $ \psi^{2}(1)\geq 0 $ we note,
\begin{eqnarray}
u_m \leq \frac{\mathfrak{n}}{2}\left( 1-\sqrt{1+\frac{4\varrho}{\tilde{\zeta}\mathfrak{n}^{2}}}+\frac{\mathfrak{K}^{2}\varrho \Gamma}{4\tilde{\zeta}r_H^{2}(3+\mathcal{A}'(1))}\right) 
\end{eqnarray}
where,
\begin{eqnarray}
\mathfrak{n}&=&\frac{4+\mathcal{A}'(1)}{3+\mathcal{A}'(1)}\nonumber\\
\Gamma &=& \frac{\varrho}{\tilde{\zeta}\mathfrak{n}^{2}\sqrt{1+\frac{4\varrho}{\tilde{\zeta}\mathfrak{n}^{2}}}}+\frac{1}{4}\left( 1- \sqrt{1+\frac{4\varrho}{\tilde{\zeta}\mathfrak{n}^{2}}}\right).
\end{eqnarray}

Finally, we match the near horizon expansion (\ref{e90}) of the scalar field ($ \psi (u) $) to that of its asymptotic expansion (\ref{E77}) which by virtue of (\ref{E75}) and (\ref{E76}) finally yields,
\begin{eqnarray}
 <\mathit{O}> &=&\frac{r_H^{2}}{2u_m f'(1)}\Upsilon(1,u_m)\xi\nonumber\\
 \Upsilon (1,u_m)&=&\left[ \left( 1-\mathcal{A}'(1)-\frac{f''(1)}{f'(1)}\right)+\frac{\zeta^{2}}{2f'(1)r_H^{4}}-\frac{2}{f'(1)}\right] (1-u_m) -2.
\end{eqnarray}

Using, (\ref{E81}) and considering the limit $ T \sim T_c $, the expression for the condensation operator ($ <\mathit{O}> $) turns out to be,
\begin{eqnarray}
 \frac{<\mathit{O}>}{T_c^{2}}&\approx & \frac{2\pi \Upsilon (1,u_m)}{\sqrt{\zeta}u_m^{3/2} f'(1)^{3/2}\sqrt{1-u_m}}\sqrt{1-\frac{T}{T_c}}\nonumber\\
& \approx & \tau (\mathfrak{K}^{2},u_m)\sqrt{1-\frac{T}{T_c}}
\end{eqnarray}
where,
\begin{eqnarray}
\frac{T_c}{\sqrt{\tilde{\varrho}}} =\left(1-\frac{\mathfrak{K}^{2}u_{m}^{2}}{16 r_H^{2}} \right)^{1/2}
\label{E84}
\end{eqnarray}
is the critical temperature corresponding to the superconducting phase transition and $ \tilde{\varrho}(=\varrho r_H) $ is the re scaled charged density for the boundary theory. Note that here we have set the constant of proportionality equal to unity.

 Before we formally conclude, a few important remarks are in order. First of all, from (\ref{E84}) one could in fact check that the value of the entity, $ \frac{T_c}{\sqrt{\tilde{\varrho}}} $ decreases with the increase in the anisotropy ($ \mathfrak{K}^{2} $). Similar effects have also been reported earlier in the context of asymptotic $AdS_{4}$ black branes. This result also supports certain experimental facts which suggest that the anisotropy eventually lowers the critical temperature in a conventional superconducting material \cite{Koga:2014hwa}. However, the crucial fact regarding our present analysis is that here we arrive at similar conclusions following analytic computations, while on the other hand, the previous analysis \cite{Koga:2014hwa} was purely based on numerical techniques. Finally, if we further demand that the gap $\left(  \frac{<\mathit{O}>}{T_c^{2}} \right)$  must increase with the anisotropy ($ \mathfrak{K}^{2} $), then one must impose certain constraint on the coefficient $ \tau $ such that it is a monotonically increasing function of the anisotropy, namely, $ \partial_{\mathfrak{K}^{2}}\tau \geq 0 $.

\section{Summary and final remarks}
We now summarise the key findings of our analysis. The upshot of the entire analysis is the following. In this paper, based on the methods proposed in \cite{Iizuka:2012wt}, we construct the anisotropic charge Lifshitz background by turning on relevant (massless) scalar deformations. We perform our analysis both in the extremal as well as in the non extremal limit of the black brane. Our analysis might be regarded as an important step for those future holographic set ups which aim to explore the physics of non Fermi liquids near its anisotropic Fermi surface. As a final application of our analysis, we probe the holographic superfluid phse of the boundary theory and explore the effects of anisotropy of the holographic $ s $- wave condensate.
\\ \\
{\bf {Acknowledgements :}}
 The author would like to acknowledge the financial support from UGC (Project No UGC/PHY/2014236).\\



\begin{thebibliography}{99}
\bibitem{Mateos:2011tv} 
  D.~Mateos and D.~Trancanelli,
  ``Thermodynamics and Instabilities of a Strongly Coupled Anisotropic Plasma,''
  JHEP {\bf 1107}, 054 (2011)
  [arXiv:1106.1637 [hep-th]].

\bibitem{Mateos:2011ix} 
  D.~Mateos and D.~Trancanelli,
  ``The anisotropic N=4 super Yang-Mills plasma and its instabilities,''
  Phys.\ Rev.\ Lett.\  {\bf 107}, 101601 (2011)
  [arXiv:1105.3472 [hep-th]].

\bibitem{Rebhan:2011vd} 
  A.~Rebhan and D.~Steineder,
  ``Violation of the Holographic Viscosity Bound in a Strongly Coupled Anisotropic Plasma,''
  Phys.\ Rev.\ Lett.\  {\bf 108}, 021601 (2012)
  [arXiv:1110.6825 [hep-th]].

\bibitem{Patino:2012py} 
  L.~Patino and D.~Trancanelli,
  ``Thermal photon production in a strongly coupled anisotropic plasma,''
  JHEP {\bf 1302}, 154 (2013)
  [arXiv:1211.2199 [hep-th]].

\bibitem{Jahnke:2013rca} 
  V.~Jahnke, A.~Luna, L.~Patiño and D.~Trancanelli,
  ``More on thermal probes of a strongly coupled anisotropic plasma,''
  JHEP {\bf 1401}, 149 (2014)
  [arXiv:1311.5513 [hep-th]].

\bibitem{Janik:2008tc} 
  R.~A.~Janik and P.~Witaszczyk,
  ``Towards the description of anisotropic plasma at strong coupling,''
  JHEP {\bf 0809}, 026 (2008)
  [arXiv:0806.2141 [hep-th]].

\bibitem{Rebhan:2012bw} 
  A.~Rebhan and D.~Steineder,
  ``Probing Two Holographic Models of Strongly Coupled Anisotropic Plasma,''
  JHEP {\bf 1208}, 020 (2012)
  [arXiv:1205.4684 [hep-th]].

\bibitem{Mamo:2012sy} 
  K.~A.~Mamo,
  ``Holographic RG flow of the shear viscosity to entropy density ratio in strongly coupled anisotropic plasma,''
  JHEP {\bf 1210}, 070 (2012)
  [arXiv:1205.1797 [hep-th]].

\bibitem{Giataganas:2012zy} 
  D.~Giataganas,
  ``Probing strongly coupled anisotropic plasma,''
  JHEP {\bf 1207}, 031 (2012)
  [arXiv:1202.4436 [hep-th]].

\bibitem{Gahramanov:2012wz} 
  I.~Gahramanov, T.~Kalaydzhyan and I.~Kirsch,
  ``Anisotropic hydrodynamics, holography and the chiral magnetic effect,''
  Phys.\ Rev.\ D {\bf 85}, 126013 (2012)
  [arXiv:1203.4259 [hep-th]].


\bibitem{Erdmenger:2012zu} 
  J.~Erdmenger, D.~Fernandez and H.~Zeller,
  ``New Transport Properties of Anisotropic Holographic Superfluids,''
  JHEP {\bf 1304}, 049 (2013)
  [arXiv:1212.4838 [hep-th]].
  
  \bibitem{Erdmenger:2010xm} 
  J.~Erdmenger, P.~Kerner and H.~Zeller,
  ``Non-universal shear viscosity from Einstein gravity,''
  Phys.\ Lett.\ B {\bf 699}, 301 (2011)
  [arXiv:1011.5912 [hep-th]].
  
  \bibitem{Erdmenger:2011tj} 
  J.~Erdmenger, P.~Kerner and H.~Zeller,
  ``Transport in Anisotropic Superfluids: A Holographic Description,''
  JHEP {\bf 1201}, 059 (2012)
  [arXiv:1110.0007 [hep-th]].
  
  \bibitem{Basu:2011tt} 
  P.~Basu and J.~H.~Oh,
  ``Analytic Approaches to Anisotropic Holographic Superfluids,''
  JHEP {\bf 1207}, 106 (2012)
  [arXiv:1109.4592 [hep-th]].
  
  \bibitem{Oh:2012zu} 
  J.~H.~Oh,
  ``Running Shear Viscosities in An-Isotropic Holographic Superfluids,''
  JHEP {\bf 1206}, 103 (2012)
  [arXiv:1201.5605 [hep-th]].
  
  \bibitem{Bhattacharyya:2014wfa} 
  A.~Bhattacharyya and D.~Roychowdhury,
  ``Viscosity bound for anisotropic superfluids in higher derivative gravity,''
  JHEP {\bf 1503}, 063 (2015)
  [arXiv:1410.3222 [hep-th]].

\bibitem{Roychowdhury:2015lma} 
  D.~Roychowdhury,
  ``Holographic charge transport in non commutative gauge theories,''
  JHEP {\bf 1507}, 121 (2015)
  [arXiv:1506.00209 [hep-th]].
  
\bibitem{Critelli:2014kra} 
  R.~Critelli, S.~I.~Finazzo, M.~Zaniboni and J.~Noronha,
  ``Anisotropic shear viscosity of a strongly coupled non-Abelian plasma from magnetic branes,''
  Phys.\ Rev.\ D {\bf 90}, no. 6, 066006 (2014)
  [arXiv:1406.6019 [hep-th]].

\bibitem{Jain:2014vka} 
  S.~Jain, N.~Kundu, K.~Sen, A.~Sinha and S.~P.~Trivedi,
  ``A Strongly Coupled Anisotropic Fluid From Dilaton Driven Holography,''
  JHEP {\bf 1501}, 005 (2015)
  [arXiv:1406.4874 [hep-th]].

\bibitem{Cheng:2014qia} 
  L.~Cheng, X.~H.~Ge and S.~J.~Sin,
  ``Anisotropic plasma at finite $U(1)$ chemical potential,''
  JHEP {\bf 1407}, 083 (2014)
  [arXiv:1404.5027 [hep-th]].

\bibitem{Koga:2014hwa} 
  J.~i.~Koga, K.~Maeda and K.~Tomoda,
  ``Holographic superconductor model in a spatially anisotropic background,''
  Phys.\ Rev.\ D {\bf 89}, no. 10, 104024 (2014)
  [arXiv:1401.6501 [hep-th]].

 \bibitem{Bai:2014poa} 
  X.~Bai, B.~H.~Lee, M.~Park and K.~Sunly,
  ``Dynamical Condensation in a Holographic Superconductor Model with Anisotropy,''
  JHEP {\bf 1409}, 054 (2014)
  [arXiv:1405.1806 [hep-th]].

\bibitem{Cheng:2014sxa} 
  L.~Cheng, X.~H.~Ge and S.~J.~Sin,
  ``Anisotropic plasma with a chemical potential and scheme-independent instabilities,''
  Phys.\ Lett.\ B {\bf 734}, 116 (2014)
  [arXiv:1404.1994 [hep-th]].


\bibitem{Iizuka:2012wt} 
  N.~Iizuka and K.~Maeda,
  ``Study of Anisotropic Black Branes in Asymptotically anti-de Sitter,''
  JHEP {\bf 1207}, 129 (2012)
  [arXiv:1204.3008 [hep-th]].


\bibitem{Kachru:2008yh} 
  S.~Kachru, X.~Liu and M.~Mulligan,
  ``Gravity duals of Lifshitz-like fixed points,''
  Phys.\ Rev.\ D {\bf 78}, 106005 (2008)
  [arXiv:0808.1725 [hep-th]].

\bibitem{Taylor:2008tg} 
  M.~Taylor,
  ``Non-relativistic holography,''
  arXiv:0812.0530 [hep-th].
  
  \bibitem{Iqbal:2011ae} 
  N.~Iqbal, H.~Liu and M.~Mezei,
  ``Lectures on holographic non-Fermi liquids and quantum phase transitions,''
  arXiv:1110.3814 [hep-th].
  
  \bibitem{Hoyos:2013eza} 
  C.~Hoyos, B.~S.~Kim and Y.~Oz,
  ``Lifshitz Hydrodynamics,''
  JHEP {\bf 1311}, 145 (2013)
  doi:10.1007/JHEP11(2013)145
  [arXiv:1304.7481 [hep-th]].
  
  \bibitem{Brynjolfsson:2010rx} 
  E.~J.~Brynjolfsson, U.~H.~Danielsson, L.~Thorlacius and T.~Zingg,
  ``Black Hole Thermodynamics and Heavy Fermion Metals,''
  JHEP {\bf 1008}, 027 (2010)
  [arXiv:1003.5361 [hep-th]].  
  
  
  \bibitem{Ogawa:2011bz} 
  N.~Ogawa, T.~Takayanagi and T.~Ugajin,
  ``Holographic Fermi Surfaces and Entanglement Entropy,''
  JHEP {\bf 1201}, 125 (2012)
  doi:10.1007/JHEP01(2012)125
  [arXiv:1111.1023 [hep-th]].
  
  \bibitem{Huijse:2011ef} 
  L.~Huijse, S.~Sachdev and B.~Swingle,
  ``Hidden Fermi surfaces in compressible states of gauge-gravity duality,''
  Phys.\ Rev.\ B {\bf 85}, 035121 (2012)
  doi:10.1103/PhysRevB.85.035121
  [arXiv:1112.0573 [cond-mat.str-el]].
  
  \bibitem{Tarrio:2011de} 
  J.~Tarrio and S.~Vandoren,
  ``Black holes and black branes in Lifshitz spacetimes,''
  JHEP {\bf 1109}, 017 (2011)
  [arXiv:1105.6335 [hep-th]].
  
  \bibitem{Hartnoll:2008vx} 
  S.~A.~Hartnoll, C.~P.~Herzog and G.~T.~Horowitz,
  ``Building a Holographic Superconductor,''
  Phys.\ Rev.\ Lett.\  {\bf 101}, 031601 (2008)
  doi:10.1103/PhysRevLett.101.031601
  [arXiv:0803.3295 [hep-th]].
  
  \bibitem{Gregory:2009fj} 
  R.~Gregory, S.~Kanno and J.~Soda,
  ``Holographic Superconductors with Higher Curvature Corrections,''
  JHEP {\bf 0910}, 010 (2009)
  doi:10.1088/1126-6708/2009/10/010
  [arXiv:0907.3203 [hep-th]].
  
  \bibitem{Roychowdhury:2012hp} 
  D.~Roychowdhury,
  ``Effect of external magnetic field on holographic superconductors in presence of nonlinear corrections,''
  Phys.\ Rev.\ D {\bf 86}, 106009 (2012)
  doi:10.1103/PhysRevD.86.106009
  [arXiv:1211.0904 [hep-th]].
  
   \end{thebibliography}
\end{document}